\documentclass[aps,prb,superscriptaddress,twocolumn,floatfix]{revtex4-1}
\usepackage{graphicx}
\usepackage{amssymb}
\usepackage{amsfonts}
\usepackage{amsmath}
\usepackage{relsize}
\usepackage{color}

\begin{document}
\affiliation{Van der Waals - Zeeman Institute, University of Amsterdam, Science Park 904, 1098 XH Amsterdam, The Netherlands}
\title{Heat capacity of type I superconductivity in the Dirac semimetal PdTe$_2$}
\author{M. V. Salis} \email{m.v.salis@uva.nl} 
\author{Y. K. Huang} 
\author{A. de Visser} \email{a.devisser@uva.nl}

\date{\today}

\begin{abstract}
Type I superconductivity has recently been reported for the Dirac semimetal PdTe$_2$ ($T_c \approx 1.6$ K) with, remarkably, multiple critical fields and a complex phase diagram. Here, measurements of the specific heat utilizing a thermal relaxation technique are presented. Conventional weak-coupling BCS superconductivity is confirmed by examining the temperature dependence of the specific heat in zero field. By probing the latent heat accompanying the superconducting transition, thermodynamic evidence for type I superconductivity is attained. The presence of the intermediate state is observed as a significant broadening of the superconducting transition onto lower temperatures at high fields as well as irreversibility in the specific heat in zero field cooled data at 8.5 mT. 

\end{abstract}

\maketitle{}

\section{Introduction}
Recently, layered transition metal dicalchogenides have sparked great interest by virtue of their exotic electronic properties, especially the possibility of realizing novel quantum states stemming from the topological non-trivial band structure as uncovered by density functional theory \cite{Soluyanov2015,Huang2016,Yan2017,Bahramy2018}. A generic coexistence of type I and type II 3-dimensional Dirac cones has been proposed to be at play in these materials \cite{Bahramy2018}. PdTe$_2$ is interesting in particular because of the appearance of superconductivity at $T_c \approx 1.6$ K \cite{Guggenheim1961,Leng2017}, as well as its classification as a type II Dirac semimetal. The latter is extracted from a combination of \textit{ab initio} electronic structure calculations and angle resolved photoemission spectroscopy \cite{Liu2015a,Fei2017,Noh2017,Bahramy2018,Clark2017}. A Dirac cone with a tilt parameter $k > 1$ breaking Lorentz invariance is the hallmark of a type II Dirac semimetal \cite{Soluyanov2015}. The Dirac point then forms the touching point of the electron and hole pockets, possibly resulting in a nearly flat band adjacent to the Fermi level. This prompts the question whether superconductivity is bolstered by the presence of the nearly flat band \cite{Rosenstein2018}. \\ 

The superconducting properties of PdTe$_2$ have been extensively investigated. Transport and magnetic measurements carried out on single crystals of PdTe$_2$ revealed the existence of bulk type I superconductivity, an uncommon feature for a binary compound \cite{Leng2017}. Dc magnetization data showed the appearance of the intermediate state, the hallmark of type I superconductivity in an applied magnetic field. This was further corroborated by the differential paramagnetic effect observed in ac magnetization measurements. A bulk critical field $B_c = 13.6$ mT was determined. A puzzling aspect is the detection of surface superconductivity with a critical field $B_c^{surf} = 34.9$ mT and a temperature dependence that does not follow the standard Saint James - de Gennes model \cite{Saint-James&deGennes1963}. This led the authors of Ref.~\onlinecite{Leng2017} to suggest surface superconductivity to have a topological nature. Moreover, an even higher critical field of 0.3 T was observed in resistance data. The theoretical possibility of type I superconductivity in PdTe$_2$ was analyzed within a microscopic pairing theory exploring the tilt parameter $k$ of the Dirac cone \cite{Shapiro2018}. The realization of type I superconductivity was established for $k = 2$.\\

Evidence for the weak-coupling Bardeen-Cooper-Schrieffer (BCS) nature of superconductivity in PdTe$_2$ was obtained through measurements of the specific heat \cite{Amit2018}, penetration depth \cite{Salis2018,Teknowijoyo2018}, scanning tunneling microscopy and spectroscopy (STM and STS) \cite{Clark2017,Das2018,Sirohi2019} and tunneling spectroscopy on side junctions \cite{Voerman2019}. Surprisingly, distinct and fairly large critical fields were observed in STM/STS measurements \cite{Das2018,Sirohi2019}, and their spatial distribution on the surface was attributed to a mixture of type I and type II superconductivity. This provided the motivation for further experimental work to unravel the nature of the superconducting phase. Additional evidence for type I superconductivity was inferred from the local electronic behavior necessary to properly analyze the magnetic penetration depth data \cite{Salis2018}. Evidence on the microscopic scale was obtained from transverse muon spin relaxation measurements in an applied magnetic field, that unambiguously demonstrated the presence of the intermediate state \cite{Leng2019}. Similarly, scanning squid magnetometry provided evidence for type I superconductivity on the macroscopic scale \cite{Garcia-Campos2020}. Finally it has been established that type I superconductivity is robust under pressure \cite{Leng2019p}.\\

Although the specific heat of PdTe$_2$ was reported before, the focus was on elucidating the symmetry of the gap structure \cite{Amit2018}. Heat capacity techniques can also be utilized to ascertain whether superconductors are type I or type II. Unlike type II superconductors, type I superconductors, when subjected to a magnetic field, will undergo a first order phase transition. This can be verified by measuring the heat capacity in a magnetic field, which involves the latent heat associated with the transition. In this case the latent heat appears as an extra contribution to the jump in the specific heat at $T_c$, such that the jump size exceeds the value in zero magnetic field. Furthermore, for type I superconducting samples that have a shape resulting in a nonzero demagnetization factor, the intermediate state emerges. The intermediate state contribution broadens the superconducting transition towards lower temperatures due to the gradual transformation of normal domains to superconducting domains. Hitherto, no thermodynamic evidence in favor of type I or type II superconductivity has been reported. This warrants a second specific heat study focusing on these aspects.\\

In this paper heat capacity measurements of PdTe$_2$ in zero and applied magnetic fields are reported. The data in field show the presence of latent heat associated with a first order transition and thus type I superconductivity. The temperature variation of the critical field, $B_c (T)$, follows the expected quadratic temperature variation up to 9.5 mT. The data at higher applied fields reveal the presence of a second, minority superconducting phase in the PdTe$_2$ crystal.

\section{Experimental}
PdTe$_2$ crystallizes in the trigonal CdI$_2$ structure (space group P$\bar{3}$m1) \cite{Thomassen1929}. The single crystal investigated in this study is taken from a batch grown with the modified Bridgman technique \cite{Lyons1976} as reported in Ref.~\onlinecite{Leng2017}. The proper 1:2 stoichiometry within the 0.5 \% experimental resolution was inferred from scanning electron microscopy (SEM) with energy dispersive X-ray (EDX) spectroscopy. Magnetization measurements showed a bulk $T_c$ of 1.64 K and $B_c = 13.6$ mT for a crystal cut from the same crystalline boule \cite{Leng2017}. The rectangular shaped single crystalline sample used in this study has sizes of $3\cdot3\cdot0.3$ mm$^3$ along the a, a$^*$ and c axes, respectively and  a mass of 39.66(2)~mg. 

Heat capacity measurements were carried out in an Oxford Instruments Heliox $^3$He refrigerator down to 0.3~K by use of the dual slope thermal relaxation calorimetry technique \cite{Stewart1983}. In this technique the sample is kept at a stable temperature $T_1$. Heat is then applied to heat the sample from $T_1$ by $\Delta T / T $ to $T_2$, which is recorded. The data recorded represents the heating curve. Subsequently the heat is removed, and the sample cools back to a stable temperature $T_1$, which is recorded as well. This represents the cooling curve. The increase in temperature $\Delta T / T \approx 1.5$~$ \%$. Both the heating and cooling curves at each temperature point are used in the analysis. The curves together form one relaxation measurement. Each specific heat data point in this study presents the average of four relaxation measurements at the same temperature, totalling eight fitted curves. 

The sample was attached to the sample platform with Apiezon N grease with the c-axis parallel to the applied magnetic field. This configuration results in a demagnetizing factor N = 0.14 \cite{Chen2002}, sufficiently large to probe the intermediate state. All measurements in a magnetic field have been carried out with the sample first cooled down in field from the normal state to the base temperature. The data points are collected by step-wise heating to the desired temperature $T_1$.

\section{Results}

\begin{figure}[b!]
\includegraphics[width = 8.6cm]{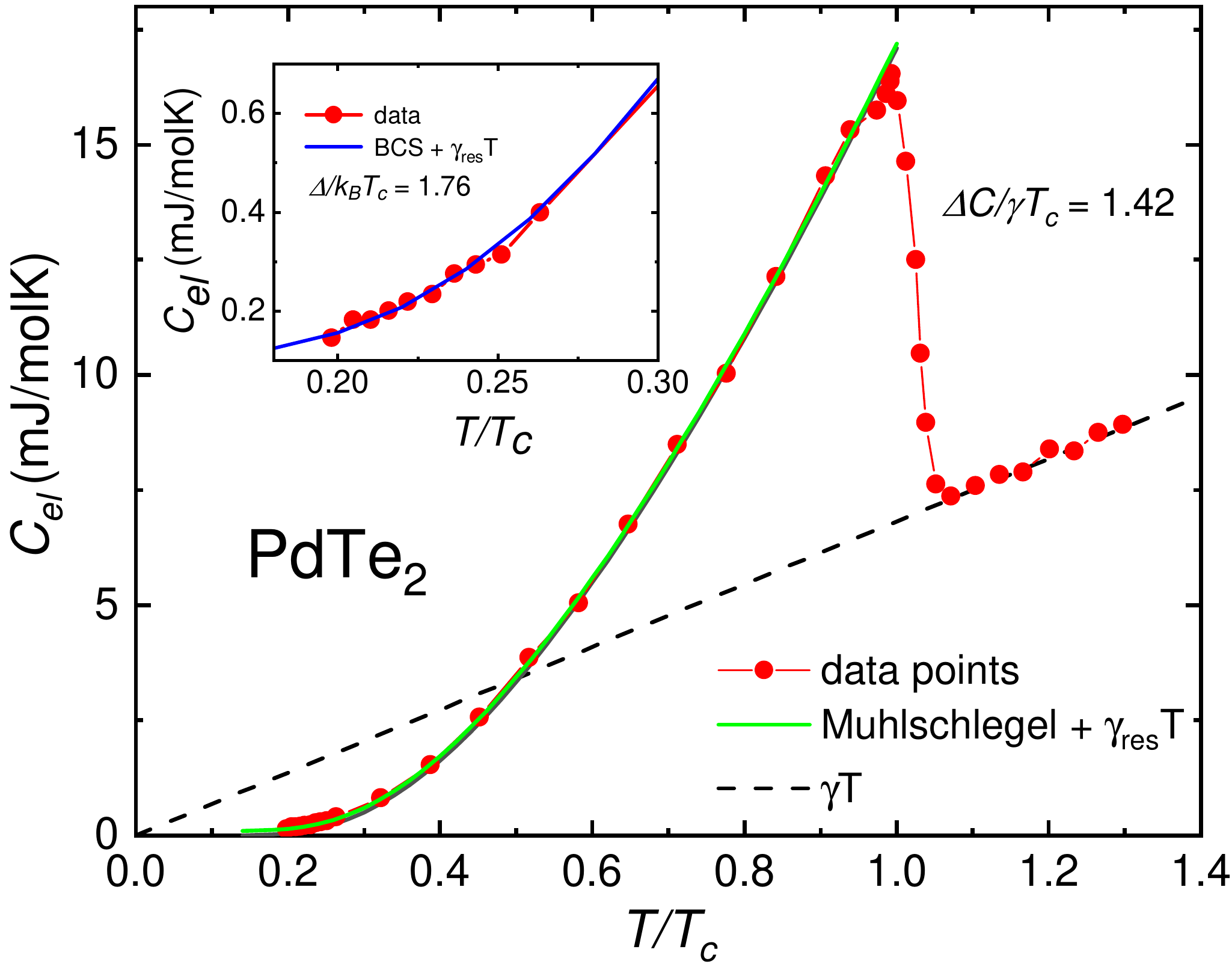}
\caption{Reduced temperature ($T/T_c$) dependence of the electronic specific heat $C_{el}$ of PdTe$_2$ in zero field. Red dots and line: experimental data; green solid line: BCS temperature dependence according to M\"{u}hlschlegel with a small residual term $\gamma_{res} T$ added; black dashed line: extrapolation to zero of the linear electronic specific heat in the normal state. The jump in the specific heat quantified with the BCS relation $\Delta C / \gamma T_c$ is equal to 1.42. Inset: Specific heat at low temperatures compared with the low temperature BCS behavior with a small residual term $\gamma_{res} T$ (see text).}\label{fig:bcs}
\end{figure}

The as-measured total specific heat, consisting of the electronic and phononic contributions, is reported in the Supplementary Material file~\cite{supp}. At low temperatures, the specific heat of a simple metal in the normal state is given by $ C = \gamma T +  \beta T^3 $, where $\gamma$ is the Sommerfeld coefficient and $\beta$ is the phononic coefficient. We have determined $\gamma$ and $\beta$ by the usual procedure~\cite{supp} and obtained values of 4.4 mJ/molK$^2$ and 0.70 mJ/molK$^4$, respectively. This $\gamma$ value compares reasonably well to the 6.0 mJ/molK$^2$ derived in previous work~\cite{Amit2018,Kubo2016}. The value $\beta$ = 0.70 mJ/molK$^4$ compares well to 0.66 mJ/molK$^4$ of the previous heat capacity study~\cite{Amit2018}. The Debye temperature $\Theta_D$ can be calculated using $\Theta_D = \bigg(\frac{S 12 \pi^4 R}{5\beta}\bigg)^{\frac{1}{3}}$, where $S$ is the number of atoms per formula unit and $R$ is the gas constant. We obtain $\Theta_D = 202$~K, which agrees well with the previously reported value of 207 K \cite{Kubo2016} and the calculated value of 211 K \cite{Amit2018}. After subtracting the phonon contribution the electronic specific heat, $C_{el}$, results.

The overal temperature variation of the electronic specific heat is presented in figure \ref{fig:bcs} in reduced temperature ($T/T_c$) with $T_c$ = 1.54~K. Here $T_c$ is taken as the temperature where $C_{el}$ has its maximum value.
\\

The jump at $T_c$ quantified with the BCS relation $\Delta C /\gamma T_c$, where $\Delta C$ is the jump in the specific heat, equals 1.42, which is close to the textbook value of 1.43, confirming the weak-coupling BCS nature of superconductivity in PdTe$_2$. The full range temperature dependence of a weak coupling BCS superconductor as tabulated by M\"{u}hlschlegel \cite{Muhlschlegel1959} is given by the green line in figure \ref{fig:bcs}. In order to better match the experimental data, a small residual linear term with $\gamma_{res} =  0.10$ mJ/molK$^2$ is added. This accounts for 2.2~$\%$ of the sample that apparently remains in the normal state. At low temperatures the superconducting specific heat is described by the relation $ C = C_n 3.5 T^{-1.5} e^{-1.76/T}$ (Ref.~\onlinecite{Muhlschlegel1959}), where $C_n$ is the specific heat of the electronic normal state at $T = T_c$. Here the BCS gap relation $\frac{\Delta}{k_B T_c} = 1.76$ is incorporated. The low temperature behavior is in full accordance with the weak coupling BCS relation as shown in the inset of figure \ref{fig:bcs}, further corroborating a conventional superconducting state in this PdTe$_2$ crystal.\\

\begin{figure}[t]
\includegraphics[width = 8.6cm]{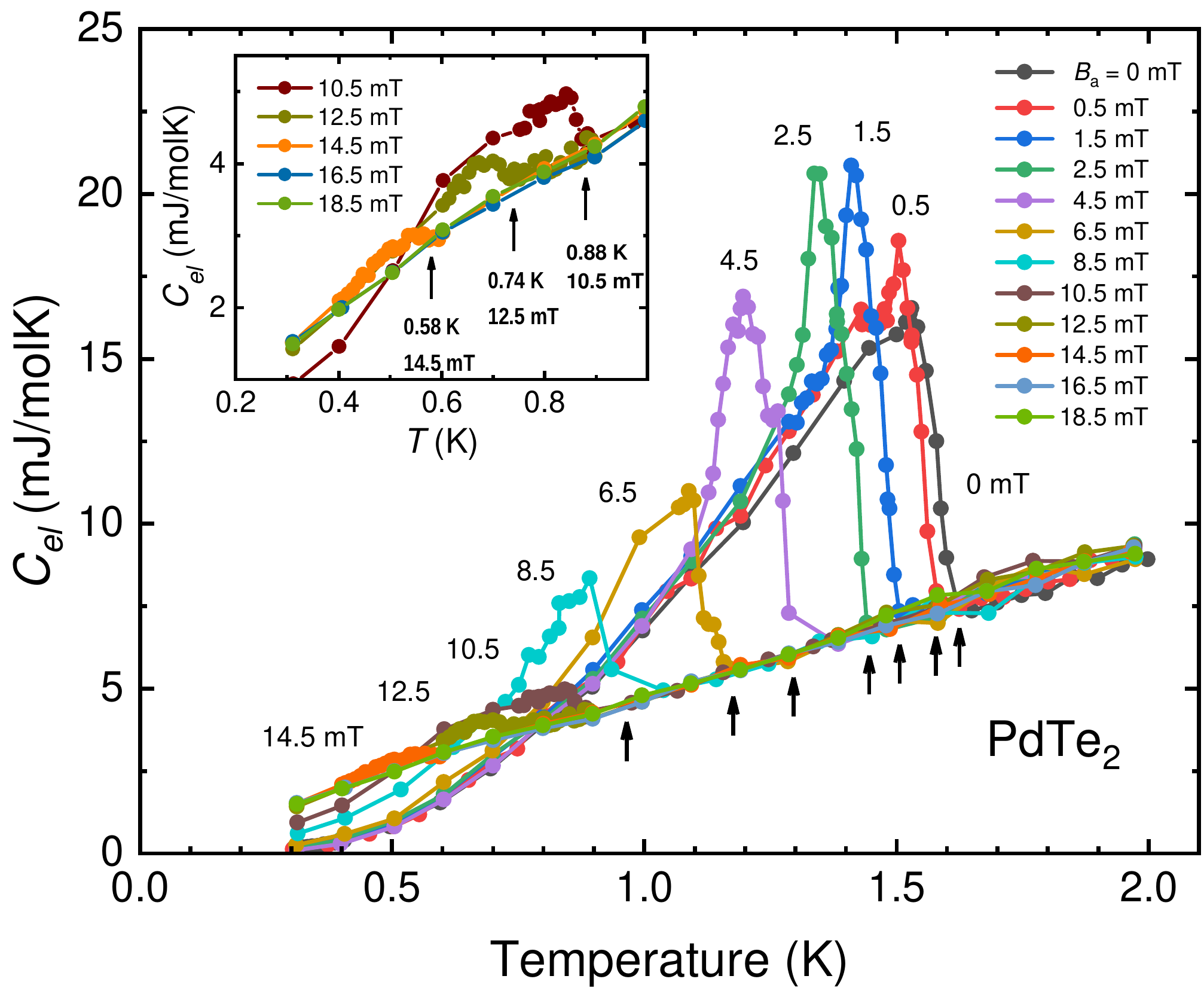}
\caption{Temperature dependence of the electronic specific heat $C_{el}$ of PdTe$_2$ in zero field and in magnetic fields up to 18.5 mT as indicated. An increase in the size of the specific heat jump at $T_c$ is observed in field. $T_c$'s are indicated by arrows.For $B \geq 10.5$ mT the jump size is strongly reduced. Inset: Zoom of the data in the low temperature range for 10.5 mT $\leq B \leq 18.5$ mT. }\label{fig:ct}
\end{figure}

\begin{figure}[h]
\includegraphics[width = 8.6cm]{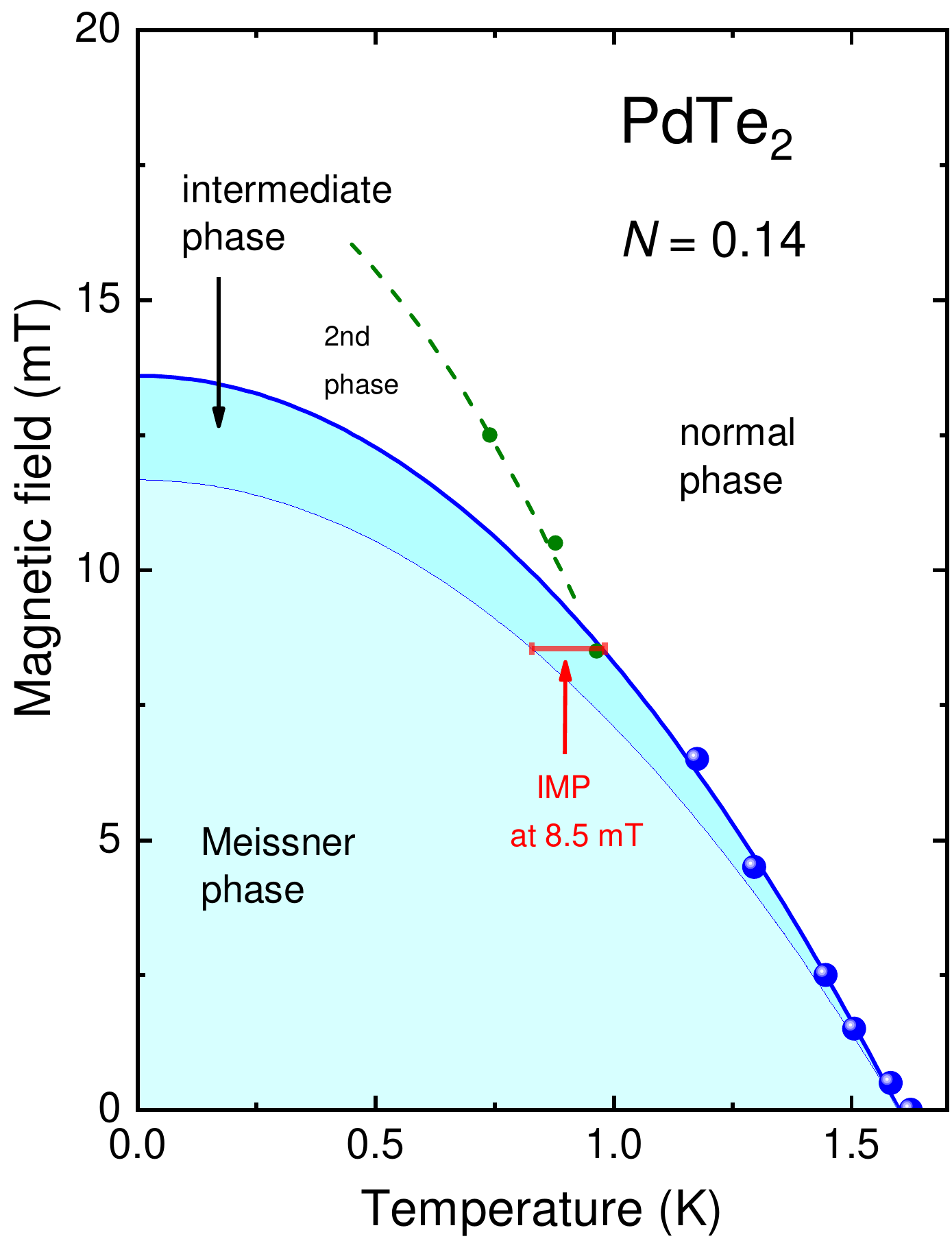}
\caption{The $B-T$ phase diagram of PdTe$_2$ obtained by plotting the onset superconducting transition temperature for different magnetic fields. Blue symbols: data points; thick solid blue line:  $B_c(T) = B_c(0) \big[1-(T/T_c)^2 \big]$ with $B_c(0) = 13.6$ mT and $T_c = 1.60$ K; thin solid blue line: Meissner-to-intermediate phase (IMP) transition line $B_{IMP}(T) = B_c(T)(1-N)$ with $N = 0.14$; green symbols: $T_c$ of a second, minority phase; dashed green line: guide to the eye;  red solid bar: temperature range of the intermediate state at 8.5 mT (see text).}\label{fig:pd}
\end{figure}

Figure \ref{fig:ct} shows the temperature dependence of the electronic specific heat, $C_{el}(T)$, in zero field and magnetic fields ranging up to 18.5 mT. The same data plotted as $C_{el}/T$ \textit{versus} $T$ are presented in Figure S2 of the Supplemental Material file~\cite{supp}. An increase in the height of the transition peak for fields up to 4.5 mT compared to the peak at 0 mT is observed. This implies extra energy is necessary to complete the transformation into the normal phase in small fields. At higher fields, especially at 6.5 mT and 8.5 mT, a broadening of the transition temperature towards lower temperatures is visible. In the experimental configuration used, the crystal has a demagnetization factor of 0.14 causing the intermediate state to form. It is likely that the superconducting transition is considerably broadened at higher fields due to the intermediate state. The region in the $B - T$ phase diagram occupied by the intermediate phase is shown in figure \ref{fig:pd}. At even higher magnetic fields, up to 16.5~mT, the transition broadens further and is no longer observed above this field. Remarkably, for $B \geq 10.5 $ mT the step size $\Delta C$ abruptly reduces.\\

In figure \ref{fig:pd} the $ B - T$ phase diagram is mapped out by tracing the onset temperatures of superconductivity in applied magnetic fields, indicated by the arrows in figure \ref{fig:ct}. In previous research \cite{Leng2017} the phase diagram for bulk superconductivity probed by different techniques was found to follow the textbook relation 
\begin{equation}
    B_c(T) = B_c(0) \big[1-(T/T_c)^2 \big],
    \label{eq:bc}
\end{equation} where $B_c(0) = 13.6$ mT and $T_c = 1.64$ K. The new data are in good agreement with the previous result with $T_c = 1.60$ K (solid blue line in figure \ref{fig:pd}). For fields $B \geq 10.5$ mT, however, we observe a somewhat higher $T_c$ than expected, which presents the onset temperature of the transition with reduced specific heat step (see the inset in figure \ref{fig:ct}). We attribute the reduced $\Delta C$ to a second, minority superconducting phase (see Discussion). The Meissner-to-intermediate phase line is given by the thin blue line. Its position is calculated by assuming that a type I superconductor is in the intermediate state for $B_c ( 1 - N) <  B_{app} < B_c$ where $B_{app}$ is the applied magnetic field and $N = 0.14$ is the demagnetization factor.\\

\begin{figure}[t]
\includegraphics[width = 8.6cm]{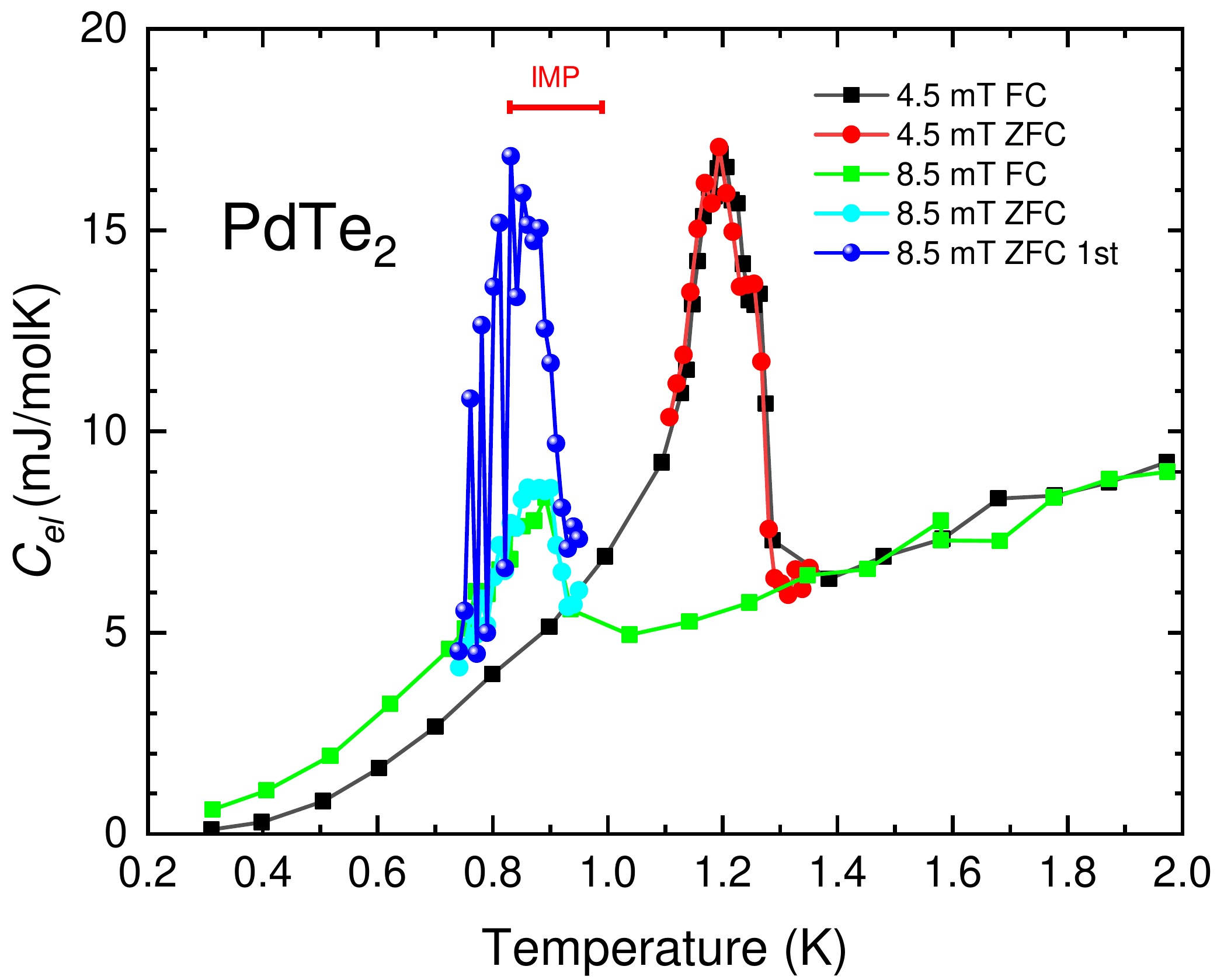}
\caption{Temperature dependence of the electronic specific heat $C_{el}$ of PdTe$_2$ measured FC and ZFC at $B$ = 4.5 and 8.5 mT. The square symbols depict FC data, whereas the round symbols depict ZFC data. No difference between FC data and ZFC data is observed at $B$ = 4.5 mT. In the ZFC data taken at 8.5 mT a large specific heat is observed, but only when derived from the heating part of the first relaxation curve (see text). No difference is observed with respect to FC data for subsequent measurements. The red bar depicts the temperature interval where the intermediate phase in 8.5 mT is expected according to figure \ref{fig:pd}.}\label{fig:zfcfc}
\end{figure}

Figure \ref{fig:zfcfc} depicts the zero field cooled (ZFC) and field cooled (FC) specific heat data as a function of temperature at 4.5 mT and 8.5 mT. All measurements here have been carried out by cooling down to base temperature either in field or without field. At the base temperature the field is applied (ZFC) or kept constant (FC). Next the sample was heated to different temperatures while keeping the field constant. The measurements carried out at 4.5 mT are given in black and red symbols, respectively, and no difference between the FC and ZFC data is found. This shows the phase transformation is the same FC and ZFC at this particular field strength. In the case of 8.5 mT, however, an odd feature is observed in the ZFC data in the temperature range 0.75-0.87 K. The heating curve of the first thermal relaxation measurement results in a much larger specific heat (see the Supplemental Material file~\cite{supp}). This is shown by the blue symbols. All subsequent data points (cyan symbols), including those derived from the cooling curve of the first relaxation measurement, fall on top of the FC data set (green symbols). This effect is only observed in the temperature range of the intermediate phase.  

\section{Discussion}
The overall temperature variation of the superconducting contribution to the specific heat is in very good agreement with the tabulated M\"{u}hlschlegel values. At the same time the low temperature data $T/T_c < 0.3$ obey the exponential expression for BCS superconductivity with $\frac{\Delta}{k_B T_c} = 1.76$ \cite{Muhlschlegel1959}. The jump size $\Delta C /\gamma T_c$ = 1.42 is conform with the weak-coupling BCS expectation of 1.43. These results compare well to previous work where a weak- to moderately coupled superconducting state and a conventional isotropic gap are reported \cite{Amit2018,Salis2018,Teknowijoyo2018,Sirohi2019,Das2018,Voerman2019}. Compared to the previous specific heat study \cite{Amit2018}, the $\gamma$ value of 4.4 mJ/molK$^2$ is nearly 20~$\%$ lower. This is possibly related to a different carrier density $n$ considering the semimetallic properties of PdTe$_2$. Differences in carrier density are also inferred from penetration depth measurements. In a previous study using single crystals from the same batch, values for the penetration depth $\lambda(0)$ were obtained that ranged from 377 nm to 482 nm\cite{Salis2018}. There $\lambda(0)$ was directly related to $n$ in an extended London model used to analyze the data where the assumption $n_s = n$ was made, with $n_s$ the superfluid density. The difference in the value for $\Delta C/\gamma T_c$ between the previous heat capacity study (1.52\cite{Amit2018}) and this work (1.42) is understood as a difference in coupling strength. This is in line with the results of penetration depth studies \cite{Salis2018,Teknowijoyo2018} where similar differences in $\Delta/k_B T_c$, ranging from 1.77 to 1.83, were found. The $\gamma$ value can be related to the critical field \cite{Poole2007}:
\begin{equation}
   \Delta C = \frac{4 B_c(0)^2}{\mu_0 T_c} = 1.43 \gamma T_c,
   \label{eq:ybc}
\end{equation}
where $\Delta C$ and $\gamma$ are per unit volume.
With the values $\gamma = 4.46$ mJ/molK$^2$, $T_c = 1.62$ K, $\Delta C / \gamma T_c = 1.42$ and the molar volume $4.34 \cdot 10^{-5}$ m$^3$/mol, we calculate $B_c(0) = 10.9$ mT. This value is smaller than the measured value $B_c(0) = 13.6$ mT. Examining the previous specific heat study\cite{Amit2018}, where $\gamma = 6.01$ mJ/mol K, $T_c = 1.8$ K and $\Delta C/ \gamma T_c$ = 1.52 were reported, eq.~\ref{eq:ybc} gives $B_c(0) = 14.6$ mT, while $B_c(0) = 19.5$ mT is the measured value. Again a similar sizeable difference is observed. As such we suspect eq.~\ref{eq:ybc} does not hold precisely for PdTe$_2$.\\

The temperature dependence of the electronic specific heat $C_{el}$ in magnetic fields shown in figure \ref{fig:ct} is consistent with that of a first order phase transition. The latent heat appearing with a first order phase transition is visible as the increased peak height in the specific heat in small fields relative to zero field. Consequently, we conclude that the type I nature of PdTe$_2$ is successfully probed via the presence of latent heat near the superconducting transition in field. Further evidence for the existence of type I superconductivity can be obtained by probing the intermediate state. In this study the sample and field geometry results in a demagnetization factor $N = 0.14$. From figure \ref{fig:pd} it is clear that for $B \geq 6.5$ mT the intermediate state spans more than 0.1 K at fixed fields, a sufficiently large interval to probe the broadening of the transition towards lower temperatures.  In the specific heat data in field the broadening towards lower temperatures is visible given the changes in the range 0.5 mT to 8.5 mT. Especially for $B \geq 6.5$ mT the broadening is very clear as the specific heat is raised considerably above the zero field value in a larger temperature range.\\

The $B_c(T)$ data points traced in figure \ref{fig:pd} closely follow the results probed by dc and ac magnetization measurements in previous work \cite{Leng2017} up to $B = 8.5$ mT. Here the phase line is the boundary for bulk superconductivity and is represented by eq. \ref{eq:bc} with $B_c(0)$ = 13.6 mT and $T_c$ = 1.60 K. However, above 8.5~mT, where $\Delta C$ is suddenly reduced, superconductivity is observed above the expected $B_c(T)$-curve. It is of importance to investigate whether this can be caused by the intermediate phase. The temperature dependence of the normal state volume fraction $F_N$ in the intermediate state in fixed fields for $B_c(1-N) < B_{app} < B_c$ is given by
\begin{equation}
F_n = \frac{1-t^2}{t_c^2 - 1} \frac{1-N}{N} + \frac{1}{N},
\label{eq:fn}
\end{equation}
where $t$ is the reduced temperature $T/T_c$, $t_c$ the reduced critical temperature $T_c(B_{app})/T_c(0)$ and $N = 0.14$ is the demagnetization factor \cite{Karl2019}. Eq. \ref{eq:fn} shows $F_n$ has a smooth temperature variation and cannot suddenly collapse. The reduced critical temperature $t_c$ in eq. \ref{eq:fn} can be rewritten using eq. \ref{eq:bc}: $t_c = \sqrt{1 - \frac{B_c(T)}{B_c(0)}}.$ From this, no sudden decrease in $F_n$ is possible as well. We therefore exclude the intermediate phase as a possible cause for the elevated $T_c$ and reduced specific heat step. A more likely explanation is that superconductivity survives in a small volume fraction ($\sim$ 10~\%) of the crystal with a slightly different PdTe$_{2+x}$ stoichiometry~\cite{Guggenheim1961}. We remark a similar additional phase line was obtained in a previous study \cite{Leng2017} by analyzing the screening signal in the ac-susceptibility for small driving fields. Since the screening signal persisted above $B_c$ it was attributed to superconductivity of the surface sheath with a critical field $B_c^{s} \approx 35$ mT.   \\

In the heating curves of the first relaxation measurements of the ZFC data detailed in figure \ref{fig:zfcfc} an increase of the specific heat at 8.5 mT appears, whereas no such increase was found at 4.5 mT. Given the temperature range in which it appears, 0.75-0.87 K, it can be attributed to the intermediate phase. We remark this range is a little lower than the expected range 0.83-0.98 K (red bar) calculated from the phase diagram in figure \ref{fig:pd}. The spatial arrangement and size of the normal and superconducting domains will depend on the field and temperature history because of pinning effects. This may cause hysteretic behavior. Such a history dependence was also reported by probing the intermediate phase in PdTe$_2$ by scanning squid magnetometry \cite{Garcia-Campos2020}. The absence of irreversibility in the relaxation curves at 4.5 mT shows the phenomenon is much weaker at this field. Moreover, the increase in specific heat in the first measurement point is more difficult to observe at 4.5 mT due to the smaller temperature range in which the intermediate phase is present. A closer examination of the irreversibility in ZFC calorimetry should be possible with ac calorimetry, as long as the change in the specific heat does not exceed the amplitude of the ac heat pulse. 

\section{Conclusion}
The temperature dependence of the specific heat of PdTe$_2$ in zero field and magnetic fields was measured in order to produce thermodynamic evidence of the type I nature of superconductivity. From the zero field data a weak-coupling BCS superconducting state is inferred conform with the literature. The data in small magnetic fields show the presence of latent heat at the superconducting transition, where the step in the specific heat $\Delta C$ exceeds the zero field value. The intermediate state was probed by (i) a significant broadening of the transition onto lower temperatures for $B > 6.5$ mT, and (ii) the appearance of irreversibility in the specific heat at 8.5 mT in ZFC data. The critical field for bulk superconductivity extracted from the data follows the standard temperature dependence with $B_c = 13.6$ mT and $T_c = 1.60$ K for $B \leq 8.5$ mT. In fields $B \geq 10.5$ mT the data reveal the presence of a second, minority phase, with a volume fraction of $\sim$ 10~\%, possibly due to  off-stoichiometric PdTe$_{2+x}$ regions.

Acknowledgement: This work is part of the Projectruimte programme with project number 680-91-109, which is financed by the Dutch Research Council (NWO).

\bibliography{References_PdTe2.bib}

\pagebreak
\widetext
\begin{center}
\textbf{\large Supplemental Material: Heat capacity of type I superconductivity in the Dirac semimetal PdTe$_2$}
\end{center}
\setcounter{equation}{0}
\setcounter{figure}{0}
\setcounter{table}{0}
\setcounter{page}{1}
\setcounter{section}{0}
\makeatletter
\renewcommand{\theequation}{S\arabic{equation}}
\renewcommand{\thefigure}{S\arabic{figure}}
\renewcommand{\bibnumfmt}[1]{[S#1]}
\renewcommand{\citenumfont}[1]{S#1}

\center{\textbf{\large 1. Measured total specific heat in zero field.}}
\begin{figure}[ht]
\includegraphics[width = 9.5cm]{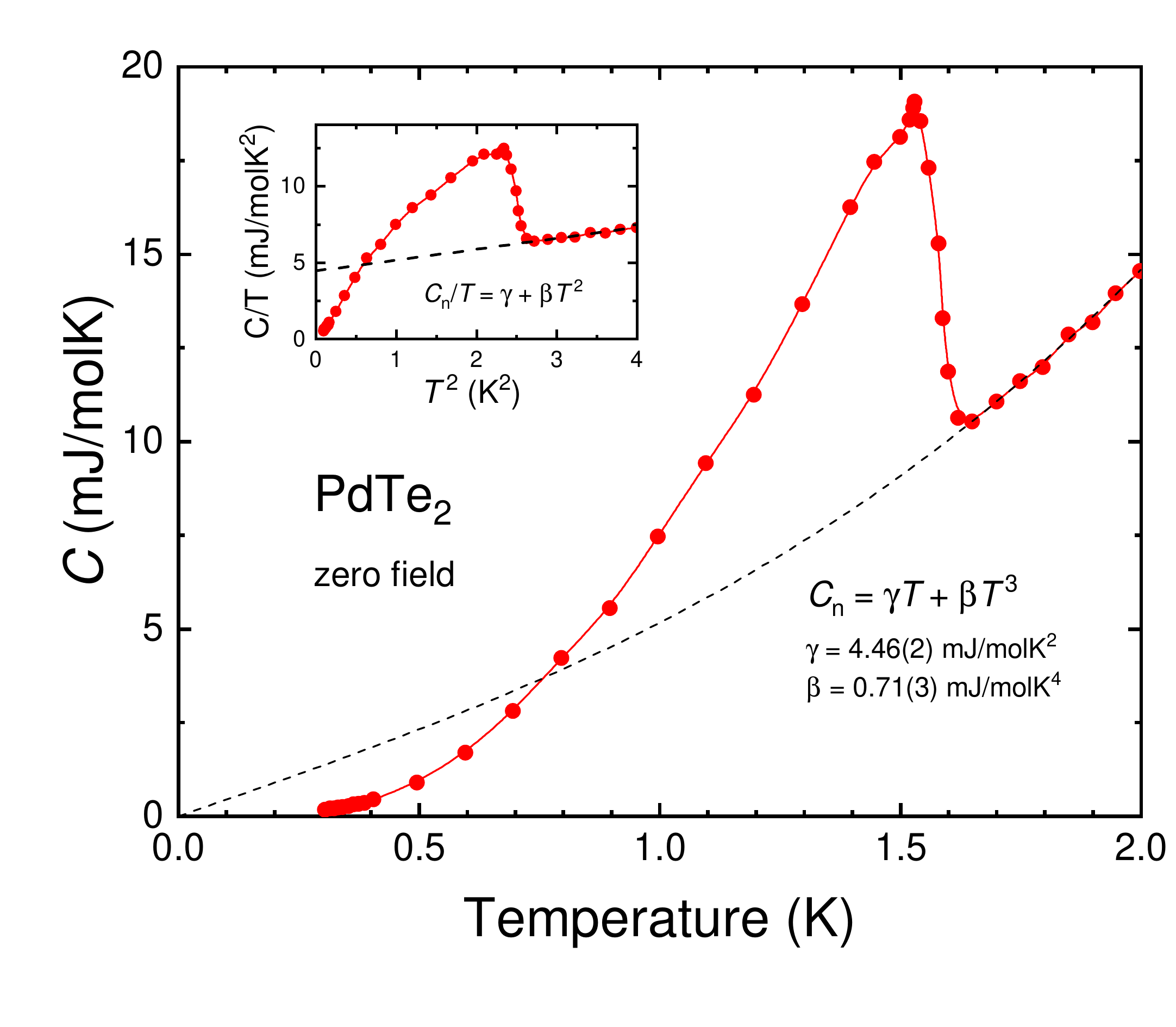}
\caption{As-measured specific heat of PdTe$_2$ in zero field in a plot of $C$ versus $T$. The dashed line shows the normal state contribution $C_n$ = $\gamma T + \beta T^3$, with the fitted values of $\gamma$ and $\beta$ listed in the graph. Inset: Same data in a plot of $C/T$ versus $T^2$.}\label{fig:cn}
\end{figure}

\center{\textbf{\large 2. Electronic specific heat in a plot of $C_{el}/T$ versus $T$ in zero and in applied magnetic fields.}}

\begin{figure}[ht]
\includegraphics[width = 9.5cm]{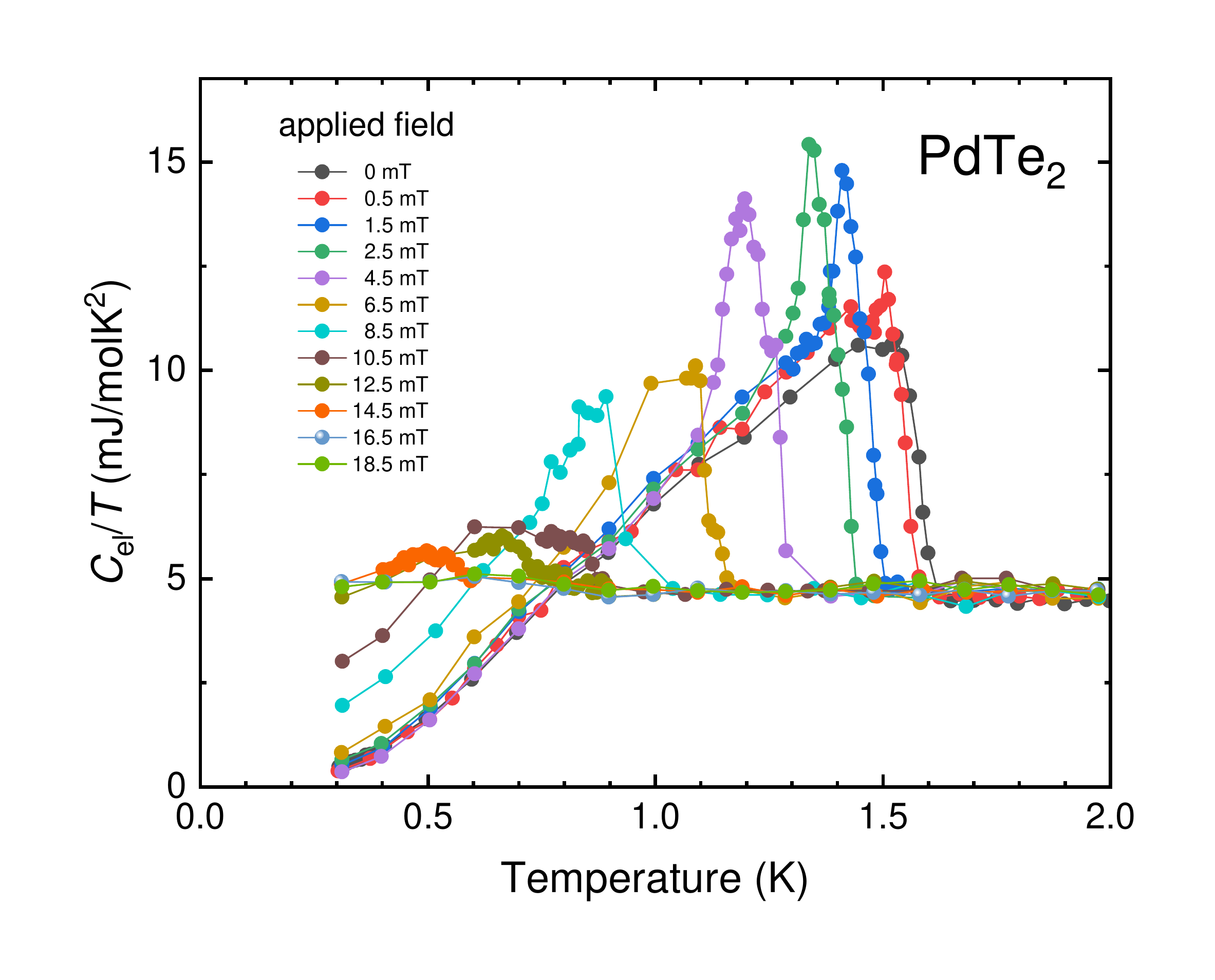}
\caption{Temperature variation of $C_{el}/T$ of PdTe$_2$ in zero field and in = magnetic fields up to 18.5 mT as indicated. The increase of the step size of the specific heat at the superconducting transition temperature measured in magnetic field is clearly observed. This is due to the first order nature of the transition which involves latent heat. For $B \geq 10.5$ mT the specific heat peak collapses.}\label{fig:ctt}
\end{figure}
\newpage

\center{\textbf{\large 3. Relaxation curves at $T = 0.8$ K and $B = 8.5$ mT.}}
\begin{figure}[ht]
\includegraphics[width = 9.5cm]{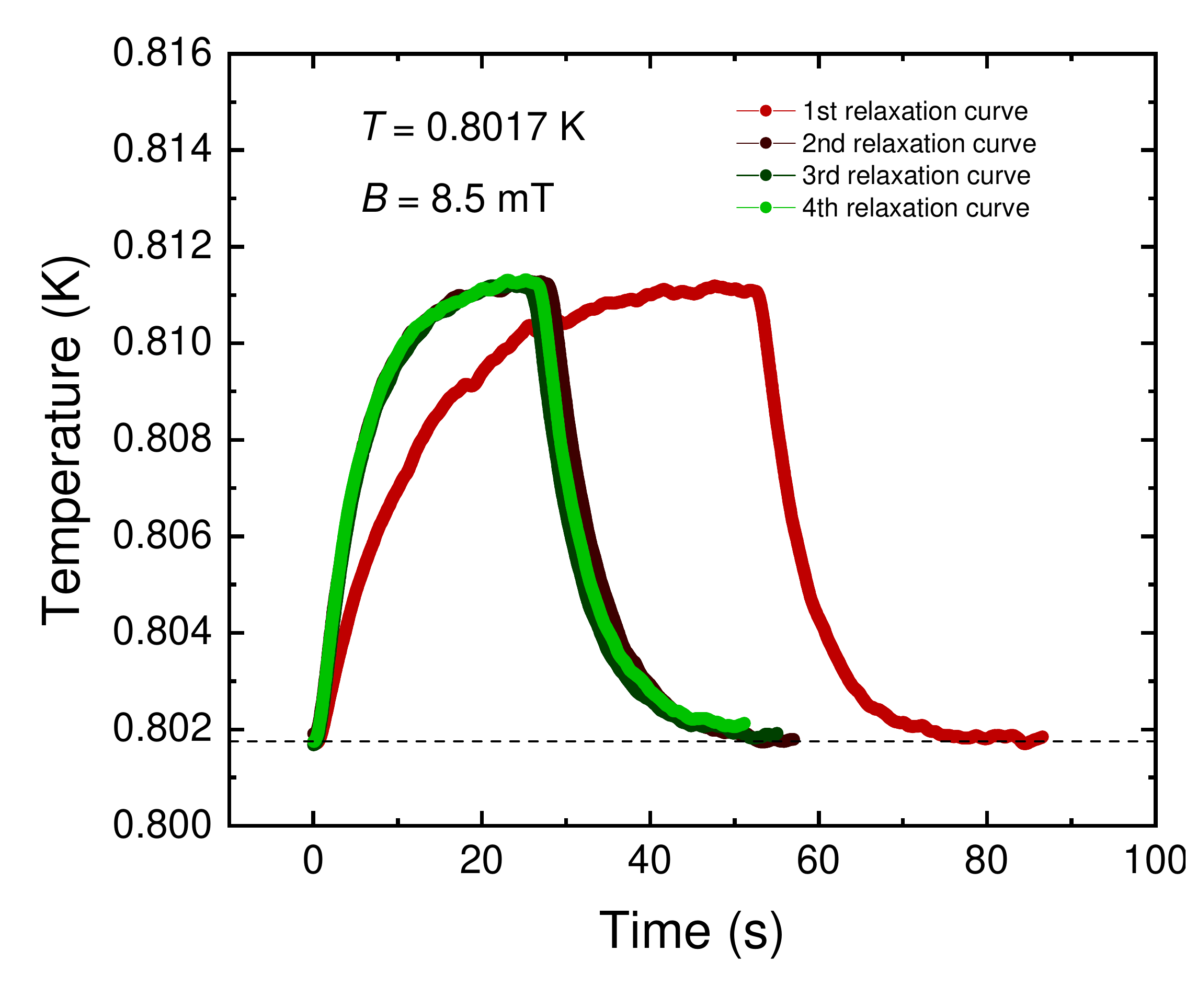}
\caption{Four subsequent relaxation curves at $T$ = 0.802 K and $B$ = 8.5 mT (applied after zero field cooling). The first heating curve (first part of the red solid line) reveals a longer relaxation time, implying a larger heat capacity. The cooling curve (second part of the red solid line) and all other heating and cooling curves show the same, shorter relaxation time. The effect is attributed to a change in the flux structure in the intermediate state in the first heating cycle.}\label{fig:hyst}
\end{figure}
\end{document}